\documentclass{iau} 
\usepackage{graphicx}
\usepackage{subcaption}
\usepackage{myshorthand}
\usepackage[round]{natbib}
\usepackage{aas_symbols}
\usepackage{aas_macros}

\title[Solar Analog Activity] 
{Evolution of Long Term Variability in Solar Analogs}

\author[Ricky Egeland]   
{Ricky Egeland$^{1,2}$, Willie Soon$^3$, Sallie Baliunas$^4$, Jeffrey
  C. Hall$^5$, Gregory W. Henry$^6$}

\affiliation{$^1$High Altitude Observatory/NCAR, 3080 Center Green Dr, Boulder CO, 80301, USA \\ 
email: {\tt egeland@ucar.edu} \\[\affilskip]
$^2$Dept. of Physics, Montana State University, P.O. Box 173840, Bozeman MT 59717, USA \\
$^3$ Harvard-Smithsonian Center for Astrophysics, Cambridge, MA 02138, USA \\
$^4$ No affiliation \\
$^5$ Lowell Observatory, 1400 West Mars Hill Road, Flagstaff, AZ 86001, USA \\
$^6$ Center of Excellence in Information Systems, Tennessee State University, 3500 John A. Merritt Blvd., Box 9501, Nashville, TN 37209, USA \\
}

\pubyear{2017}
\volume{328}  
\jname{Living around Active Stars}
\editors{D. Nandi, A. Valio, P. Petit}
\begin{document}

\maketitle

\begin{abstract}
Earth is the only planet known to harbor life, therefore we may
speculate on how the nature of the Sun-Earth interaction is relevant
to life on Earth, and how the behavior of other stars may influence
the development of life on their planetary systems.  We study the
long-term variability of a sample of five solar analog stars using
composite chromospheric activity records up to 50 years in length and
synoptic visible-band photometry about 20 years long.  This sample
covers a large range of stellar ages which we use to represent the
evolution in activity for solar mass stars.  We find that young, fast
rotators have an amplitude of variability many times that of the solar
cycle, while old, slow rotators have very little variability.  We
discuss the possible impacts of this variability on young Earth and
exoplanet climates.
\keywords{Sun Activity Chromosphere Variability Climate Habitability}
\end{abstract}

\firstsection 
\section{Introduction}

The cyclic variability of the solar sunspot count was noted by
\citet{Schwabe:1844}, but observations of surface \emph{activity} in
Sun-like stars came more than a century later.  \citet{Wilson:1968}
introduced the Mount Wilson Observatory (MWO) HK Project, which began
synoptic monitoring of the emission
in the cores of Fraunhofer H (3968.47 \AA{}) and K (3933.66 \AA{}) K
lines for a sample of Sun-like (F, G, and K-type) stars.  Formed by singly-ionized calcium, these lines have a
reversal feature for which the central emission has long been known to
correlate with regions of strong magnetic field on the Sun
\citep{Leighton:1959,Linsky:1970}.  Using Ca \II{} H \& K as a proxy
for magnetic activity on stars, \citet{Wilson:1978} presented
observations of 91 main-sequence stars, showing that they do in fact
vary, and that several of the stars appeared to have completed a cycle
in HK flux variations.  \citet{Baliunas:1995} summarized $\sim$25 years
of synoptic observations for 112 stars and conclusively showed the
existence of cyclic variability, as well as other patterns of
variability.  More than half the sample showed either erratic
variability, long-term trends, or flat activity that may be analogous
to the solar Maunder Minimum, a long period of subdued solar activity
from about 1650--1700 \citep{Eddy:1976}.

\citet{Wilson:1968} discussed the difficulty of detecting variability
in broad-band visible observations, estimating a 0.001 magnitude
(1 millimagnitude (mmag); $\approx$0.1\%) change in solar luminosity due to the passage of
spots covering about 1400 millionths of the solar surface.  This is
comparable to the later measurements of the average cyclic variation
in the total solar irradiance (TSI) from the Solar Maximum Mission
\citep{Willson:1991}.  The challenge of measuring visible-band
variability in Sun like (FGK-type) stars was taken up by researchers
at Lowell Observatory, who used differential photometry of the Str\"{o}mgren $b$
and $y$ bands to achieve the required precision
\citep{Lockwood:1997}.  They found short-term (inter-year) and
long-term (year-to-year) rms amplitudes ranging from 0.002 mag (0.2\%)
to 0.07 mag (7\%) for about 41 program stars.  Overlap with MWO
targets allowed the comparison of photometric variability in the
combined bandpass ($(b+y)/2$)
to Ca \II{} H \& K activity expressed with the $\RpHK$ index, the
ratio of HK flux to the bolometric luminosity.
\citet{Lockwood:1997} generally found that active stars (high $\RpHK$)
have larger rms photometric variability, and \citet{Radick:1998} found a
power law relationship between the two quantities.  Furthermore,
\citet{Radick:1998} found that stars were either
\emph{faculae-dominated}, with a positive correlation between
brightness and H \& K activity, or \emph{spot-dominated}, with a
negative correlation.  The terminology here refers to the dominant
features contributing to visible-band brightness variations.  The
faculae are the photospheric counterpart to the \emph{plage} in the
chromosphere, which are bright features in Ca \II{} H \& K, while
spots are dark features in both H \& K emission and visible
bandpasses.

Stars like the Sun emit most of their flux in the visible
spectrum, and for a planet with an atmosphere like the Earth's, the
majority of the radiant energy reaching the surface likewise comes in
the visible.  The $\sim$0.1\% variability in TSI from the present day
Sun is thought to be of little consequence to the globally averaged
Earth temperature \citep{Stocker:2013}, however this may not have always
been the case.  The climate impact of the Maunder Minimum period, and
its coincidence with the Medieval Little Ice Age are actively
debated, however interpretations are crucially dependent on the use of proxy records to
extrapolate the present TSI into the Maunder Minimum period
\citep[e.g.][]{Kopp:2014,Solanki:2013}.  The stellar studies of \citet{Radick:1998}
and \citet{Lockwood:2007} show a clear relationship between visible
band variability and Ca \II{} H \& K activity, and furthermore it has
long been known that stellar activity decreases with age as a star
loses angular momentum
\citep{Skumanich:1972,Noyes:1984,Barnes:2007}.  We
therefore ask the question, ``how has solar variability impacted
Earth's climate on stellar evolution (billion year) timescales?'', and
the related question, ``how might stellar variability affect exoplanet
climates?''

Clearly the most important impact of stellar evolution 
on planetary climate is the total flux reaching the top of the atmosphere.
According to standard solar evolution models, the luminosity of the
Sun has been steadily increasing from an initial value of $\sim$70\%
the present-day luminosity when hydrogen burning began
$\sim$4.6 billion years ago \citep[e.g.][]{Gough:1981}.  Because of the lower luminosity, from
first-order calculations we would expect the mean temperature of the
Earth to be below the freezing point of water, which is in
contradiction to geological evidence for wet conditions and the development of life on
Earth 3--4 billion years ago \citep{Sagan:1972}.  This problem is known as the ``Faint
Young Sun Paradox,'' which was discussed by Dr. Martens at this
symposium.  In this contribution, we shall ignore the
mean luminosity and consider the climate impact of decadal scale
variability from younger, more active stars.


\section{Long-term Variability of Solar Analogs}

\begin{figure}[ht!]
  \centering
  \includegraphics[width=\textwidth]{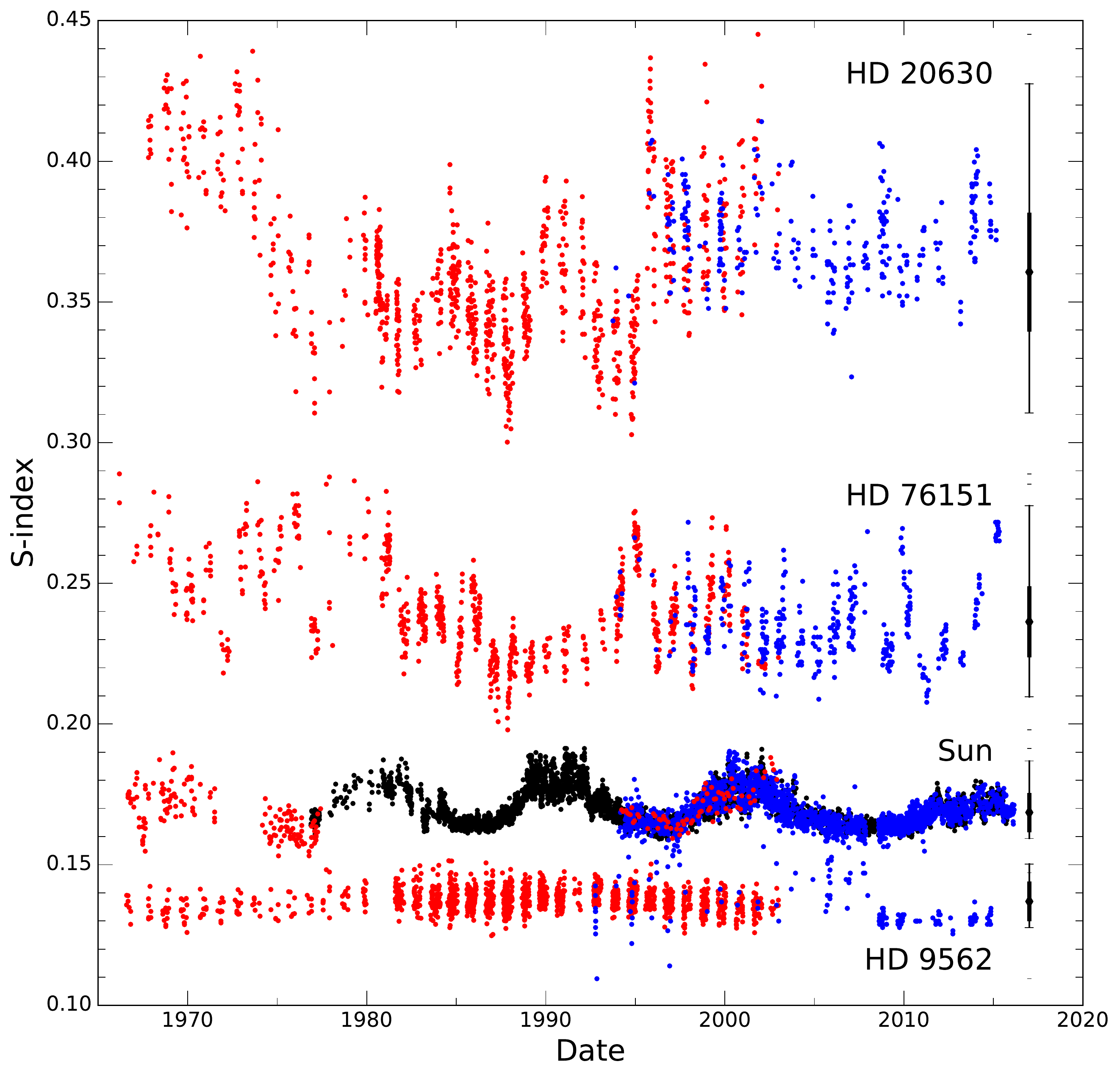}
  \caption{Calibrated composite MWO (red) + SSS (blue) time series for
    the Sun and three solar analogs.  The relative offsets of each
    time series are real.  Data from the Sun are those described in
    \citet{Egeland:2017}.  The black bar symbol on the right
    of each time series indicates four quantities: (1) the middle
    diamond is at the median $S$ for the complete time series, (2) the
    thin capped bar indicates the location of the 1st and 99th
    percentile of the data (3) the small dashes indicate the minimum
    and maximum points and (4) the thick bar is the median seasonal
    inner-98\% amplitude.}
  \label{fig:stack}
\end{figure}

To begin to address the questions of the relationship between stellar
variability and planetary climate, we look at a sample of five
solar-analog stars that may represent the behavior of the Sun at
various points in the history of the solar system, as shown in Table
\ref{tab:sample}.  This sample is drawn from a larger sample of 26
solar-analog stars with Ca \II{} H \& K observational records up to 50
years in length.  These long records are obtained by combining
observations from the MWO HK Project (1966--2003) and the Lowell
Observatory SSS (1994--present).  Some initial results from this study
were presented in \citet{Egeland:2016:cs19} and
\citet{Egeland:2016:cs19poster}.  A similarly long Sun-as-a-star Ca
\II{} H \& K record was developed in \citet{Egeland:2017}, which
accurately placed the long NSO Sacramento Peak Ca \II{} K-line record on the
$S$-index scale using coincident observations of the Moon from the MWO
HKP-2 instrument.  Figure \ref{fig:stack} shows the solar $S$-index
record and three other stars from our sample on the same scale,
illustrating the range of mean activity levels and amplitudes.  The
youngest, most active star in our sample is HD 20630 ($\kappa^1$
Ceti), which was discussed at this symposium by Dr. Dias do
Nascimento, Jr.  Not shown are HD 30495 and HD 146233 (18 Sco), the
former which is studied in detail in \cite{Egeland:2015}, and the
latter which is a solar twin \citep{PortoDeMello:1997,Melendez:2014}
and has a mean activity and amplitude very similar to the Sun
\citep{Hall:2007b,Egeland:2017}.

Table \ref{tab:sample} shows the properties of the sample.  All the
stars are within 2\% of the solar effective temperature.  All but HD
9562 lie very near to the 1 $M_\Sun$ evolutionary track, and therefore
approximate the Sun at different points in its lifetime from an age of
0.5 Gyr to the present Sun.  HD 9562 is a subgiant which has cooled
into the temperature range of our ``solar analog'' definition, and is
more massive than the Sun and the other stars in the
sample.  Using $\log g = 3.99 \pm 0.01$ from \citet{Lee:2011} and the
radius from Table \ref{tab:sample} we obtain $M/\Mnom = 1.24 \pm
0.05$.  Its slow rotation and increased radius are representative of
a future Sun, however the Sun is expected to have a lower
surface temperature when it similarly expands
\citep[see][]{Bressan:2012}.

\begin{table}
  \begin{center}
  \caption{Stellar Properties \& Variability}
  \label{tab:sample}
  \small
  \begin{tabular}{lcccccc}\hline
  {\bf Quantity} & {\bf HD 20630} & {\bf HD 30495} & {\bf HD 76151} & {\bf HD 146233} & {\bf Sun} & {\bf HD 9562} \\\hline
  $M_V$         & 5.04   & 4.87  & 4.81  & 4.79  & 4.82 & 3.41 \\
  $T/\Tnom$     & 0.99   & 1.00  & 0.98  & 1.00  & 1 & 1.01 \\ 
  $R/\Rnom$     & 0.93   & 0.97  & 1.05  & 1.02  & 1 & 1.85 \\ 
  $L/\Lnom$     & 0.83   & 0.95  & 1.03  & 1.03  & 1 & 3.62 \\
  \lbrack Fe/H\rbrack & 0.00   & $-$0.08 & $-$0.04 & $-$0.02 & 0 & +0.13 \\
  $P_\rot$ [d] & 9.2    & 11.36 & 15.0  & 22.7  & 26.09 & 29.0 \\
  Age [Gyr]    & $0.5 \pm 0.1$    & $1.0 \pm 0.1$  & $1.4 \pm 0.2$   & $3.66^{+0.44}_{-0.50}$  & 4.57 & $3.4^{+1.7}_{-0.2}$  \\\hline
  $\median{S}$ & 0.3606 & 0.3020 & 0.2363 & 0.1703 & 0.1686 & 0.1369 \\
  $A_{S,98\%}$ & 0.1169 & 0.0708 & 0.0679 & 0.0414 & 0.0275 & 0.0226 \\  
  $A_{S,s}$    & 0.0902 & 0.0502 & 0.0576 & 0.0313 & 0.0203 & 0.0159 \\  
  $A_{by,s}$ [mmag]   &   30.1 & 21.5 & 7.9 & 1.3 & 1.5$^\star$ & 1.8 \\\hline
  \end{tabular}
  \end{center}
  \vspace{0.5em}

  {\it Notes:}
  Stellar $M_V$, $T_\eff$, and [Fe/H] are from the Geneva-Copenhagen
  survey \citep{Holmberg:2009}.  Stellar luminosities are derived
  using the empirical bolometric correction of \citep{Flower:1996},
  and radii follow from the Stephan-Boltzmann Law.  $T$, $R$, and $L$
  are given in solar units using the IAU 2015 resolution B2 nominal
  values \citep{Prsa:2016} and have an uncertainty of 1--2\%.  Rotation periods are from (in order)
  \citet{Gaidos:2000}, \citet{Egeland:2015} (E15),
  \citet{Donahue:1997b}, \citet{Petit:2008}, \citet{Donahue:1996},
  \citet{Baliunas:1996}.  Ages are from (in order) \citet{Barnes:2007}
  (B07), E15, B07, \citet{Li:2012}, \citet{Bouvier:2008},
  \citet{Holmberg:2009}.  The solar $(b+y)/2$ amplitude is estimated
  in \cite{Lockwood:2007} by applying a scaling factor to the TSI
  record.
\end{table}

We have computed two estimates of the amplitude of variability in the
$S$-index records of this sample.  The first is the inter-98\% range,
$A_{S,98\%}$, taken as the difference between the top and bottom 1\%
of the $\sim$50 year time series.  The thin bars in Figure
\ref{fig:stack} demonstrate this estimate of amplitude.  We also
computed full range of the timeseries of seasonal median activity,
$A_{S, s}$.  Both of these measures of amplitude increase
monotonically with median activity, $\median{S}$, but decrease with
rotation period, $P_\rot$.  In fact, from the larger sample of 26
stars we find good linear relationships between median activity and
the amplitude, while the relationship with rotation period has
significant scatter (Egeland et al. 2017, in preparation).  On the
Sun, the $S$-index is a proxy for surface magnetic flux
\citep[e.g.][]{Harvey:1999,Pevtsov:2016}.  From Table \ref{tab:sample}
we conclude that the younger Sun had not only higher mean levels of
surface flux, but also significantly larger variation in surface flux
over decadal timescales.  The most active star in our sample, HD
20630 ($\kappa^1$ Ceti), has an amplitude of $S$-index variability over four times the
solar amplitude.  The variability is quite erratic, as can be seen in
Figure \ref{fig:stack}, but a period of reduced activity persists for
about two decades from 1975 to 1995.  HD 30495 varies by 2.5 times the
solar amplitude, though it appears to have a semi-regular cycle with a
period of $\sim$12 yr \citep{Egeland:2015}.  HD 76151 is also varying
with about 2.5 times the solar amplitude, but quite erratically.  HD
146233 (18 Sco) varies with a 50\% larger amplitude than the Sun in a
cyclic fashion \citep{Hall:2007b}, while the subgiant HD 9562 has an
amplitude about 20\% smaller.

\begin{figure}[ht!]
  \centering
  \begin{subfigure}{0.49\linewidth}
    \centering
    \includegraphics[width=\linewidth]{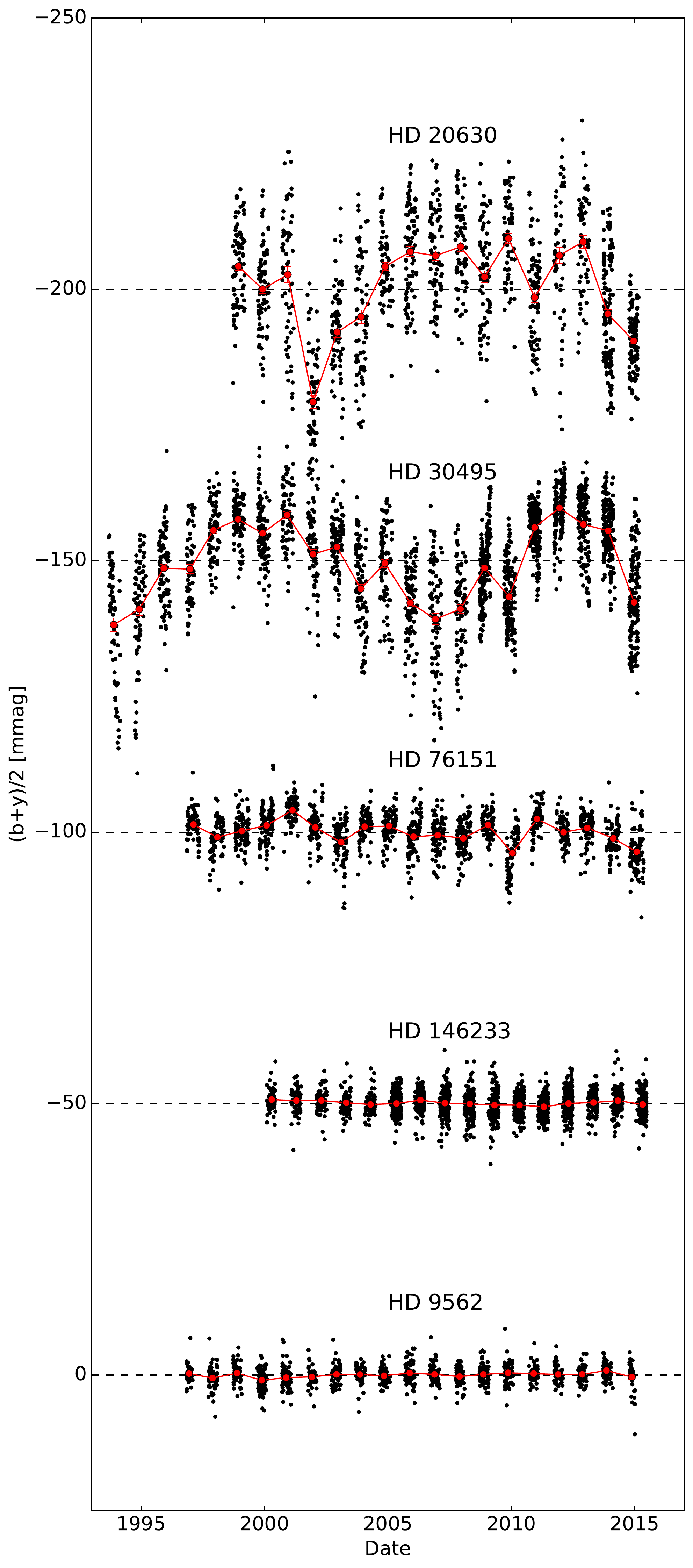}
    \caption{}
  \end{subfigure}
  \begin{subfigure}{0.49\linewidth}
    \centering
    \includegraphics[width=\linewidth]{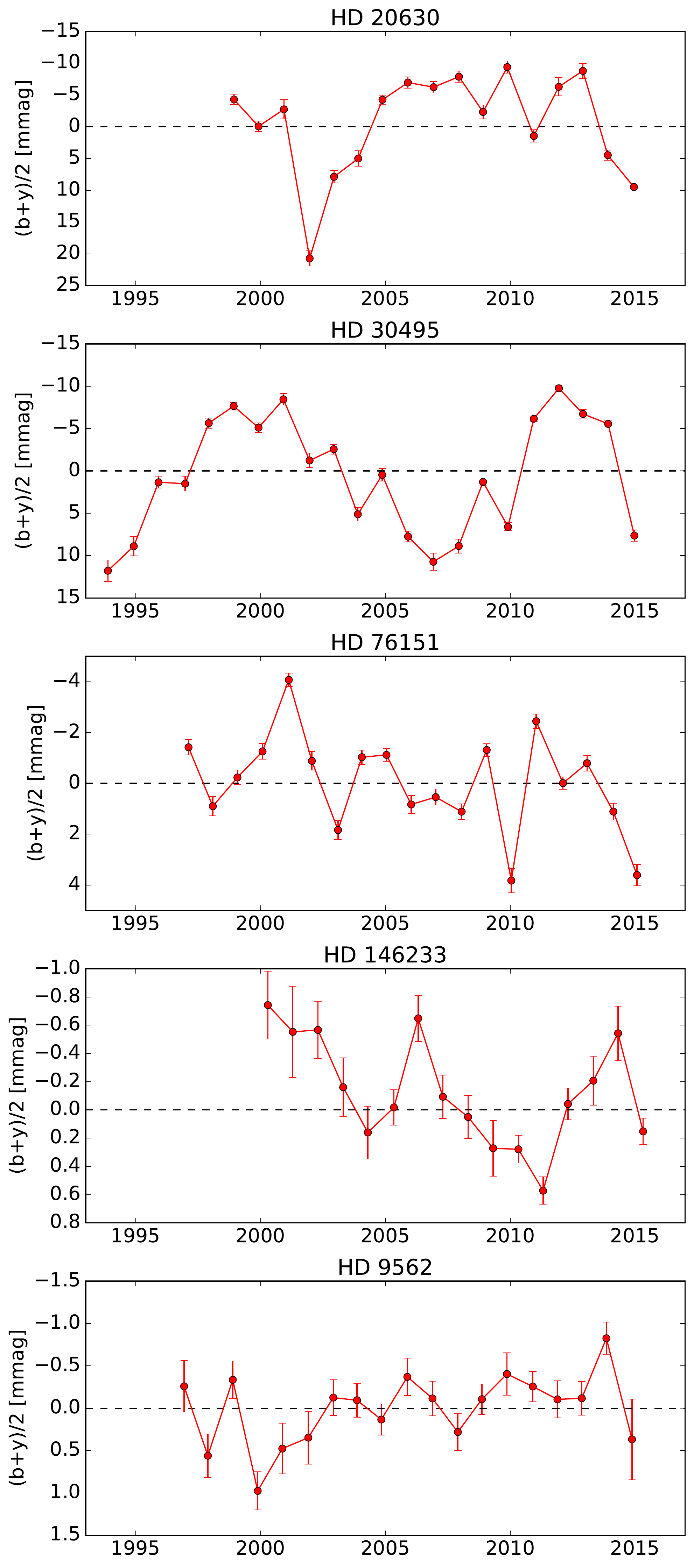}
    \caption{}
  \end{subfigure}
  \caption{Variability in the combined Str\"{o}mgren $(b + y)/2$
    bandpass from the Fairborn Observatory APT differential
    photometry.  Note that the y-axis is reversed so that higher
    points represent higher brightness.  Panel (a) shows the nightly
    measurements for each star on the same scale, with the arbitrary
    mean value shifted in increments of 50 mmag.  Seasonal means are
    shown with red points, and are shown again with a smaller scale in
    panel (b).  Error bars indicate the uncertainty of the seasonal
    mean.}
  \label{fig:bystack}
\end{figure}

Figure \ref{fig:bystack} shows the variability of our sample in
Str\"{o}mgren $(b+y)/2$ from the Fairborn Observatory Automated
Photometric Telescopes \citep[APT;][]{Henry:1995}.  The photometric
brightness is measured in millimagnitudes (mmag).
In these visible bandpasses the range of variability across the sample
is even more pronounced.  The full range of the seasonal means,
$A_{by,s}$, is shown in Table \ref{tab:sample}.  The Sun's variability
in $b+y$ is not well known, but \citet{Lockwood:2007} estimate it from
the TSI variations and a blackbody approximation of spectral
irradiance.\footnote{The SORCE SIM instrument measured a
  \emph{negative} correlation between solar activity and irradiance
  from 400--691 nm, which covers the $b$ and $y$ bands
  \citep{Harder:2009}.  This surprising result remains controversial
  \citep[e.g.][]{Yeo:2014}, and new observations will be required to
  settle the matter.  \citet{Judge:2015} proposes to place
  well-characterized reflector into geosynchronous orbit from which a
  very long, stable timeseries of spectral irradiance using
  differential photometry could be obtained.}  Note that the solar
twin 18 Sco has a similar amplitude of variability to the solar
estimate.  We find that for HD 20630 the amplitude of variability in
the visible varies by about \emph{twenty times} the estimated solar
value.  HD 30495 varies by about 14 times the solar amplitude, and HD
76151 about 5 times.  The flat-activity subgiant 9562 varies slightly
more in the visible than the Sun and 18 Sco.

\section{Consequences for Planetary Climate}

What would be the impact on Earth climate if Sun were to vary by
\emph{twenty times} its present value in the visible, as does the
young solar analog HD 20630 ($\kappa^1$ Ceti)?  If HD 20630 represents
the Sun at an age of $\sim$500 Myr, then this greatly enhanced
variability took place $\sim$4.1 Gya, at a time when life may have
been forming on Earth \citep{Bell:2015}.  Did the enhanced variability
play a role in the development of life on Earth?  Is such stellar
variability a significant factor in determining the habitability of
exoplanets?

To begin to address these questions, we consider the Earth climate
study of \citet{Meehl:2013}, who asked whether a future Maunder
Minimum-type event might significantly slow global warming.
\citet{Meehl:2013} used the Whole Atmosphere Community Climate Model
(WACCM) and modified the solar TSI input to include a step-function 0.25\% decrease
lasting 50 years.  The model produced an immediate response in globally averaged temperature
to this small decrease in TSI compared
to the baseline case with no prolonged TSI decrease, reducing global
temperature by several tenths of a degree centigrade.  However,
following the period of decreased TSI the warming trend resumed and caught
up with the baseline case.  Thus, \citet{Meehl:2013} concludes that a
future Maunder Minimum-like event could slow down, but not stop the
global warming trend.

For our purposes, the significant point is that the global temperature
registered an immediate response to the small 0.25\% decrease in TSI.
When the Sun was like HD 20630, it may have produced a much larger
variations of the order 1-2\%.  What would solar variability such as
this entail for the primitive Earth surface atmosphere and oceans,
which were significantly different not only in composition (much less
oxygen), but in structure (continental shifts)?  More detailed
theoretical work is required to determine the importance of such
enhanced stellar variability on ancient Earth and exoplanet climate.

\vspace{2em}
R.E. thanks the organizers for the invitation to this symposium and the
travel funding provided by the AAS-SPD Thomas Metcalf award.
R.E. is supported by the Newkirk Fellowship at the NCAR High Altitude
Observatory.

\bibliographystyle{abbrvnat}
\bibliography{iaus328proc}

\begin{thebibliography}{46}
\providecommand{\natexlab}[1]{#1}
\providecommand{\url}[1]{\texttt{#1}}
\expandafter\ifx\csname urlstyle\endcsname\relax
  \providecommand{\doi}[1]{doi: #1}\else
  \providecommand{\doi}{doi: \begingroup \urlstyle{rm}\Url}\fi

\bibitem[{Baliunas} et~al.(1995){Baliunas}, {Donahue}, {Soon}, {Horne},
  {Frazer}, {Woodard-Eklund}, {Bradford}, {Rao}, {Wilson}, {Zhang}, {Bennett},
  {Briggs}, {Carroll}, {Duncan}, {Figueroa}, {Lanning}, {Misch}, {Mueller},
  {Noyes}, {Poppe}, {Porter}, {Robinson}, {Russell}, {Shelton}, {Soyumer},
  {Vaughan}, and {Whitney}]{Baliunas:1995}
S.~L. {Baliunas}, R.~A. {Donahue}, W.~H. {Soon}, J.~H. {Horne}, J.~{Frazer},
  L.~{Woodard-Eklund}, M.~{Bradford}, L.~M. {Rao}, O.~C. {Wilson}, Q.~{Zhang},
  W.~{Bennett}, J.~{Briggs}, S.~M. {Carroll}, D.~K. {Duncan}, D.~{Figueroa},
  H.~H. {Lanning}, T.~{Misch}, J.~{Mueller}, R.~W. {Noyes}, D.~{Poppe}, A.~C.
  {Porter}, C.~R. {Robinson}, J.~{Russell}, J.~C. {Shelton}, T.~{Soyumer},
  A.~H. {Vaughan}, and J.~H. {Whitney}.
\newblock {Chromospheric variations in main-sequence stars}.
\newblock \emph{\apj}, 438:\penalty0 269--287, Jan. 1995.
\newblock \doi{10.1086/175072}.

\bibitem[{Baliunas} et~al.(1996){Baliunas}, {Nesme-Ribes}, {Sokoloff}, and
  {Soon}]{Baliunas:1996}
S.~L. {Baliunas}, E.~{Nesme-Ribes}, D.~{Sokoloff}, and W.~H. {Soon}.
\newblock {A Dynamo Interpretation of Stellar Activity Cycles}.
\newblock \emph{\apj}, 460:\penalty0 848, Apr. 1996.
\newblock \doi{10.1086/177014}.

\bibitem[{Barnes}(2007)]{Barnes:2007}
S.~A. {Barnes}.
\newblock {Ages for Illustrative Field Stars Using Gyrochronology: Viability,
  Limitations, and Errors}.
\newblock \emph{\apj}, 669:\penalty0 1167--1189, Nov. 2007.
\newblock \doi{10.1086/519295}.

\bibitem[Bell et~al.(2015)Bell, Boehnke, Harrison, and Mao]{Bell:2015}
E.~A. Bell, P.~Boehnke, T.~M. Harrison, and W.~L. Mao.
\newblock Potentially biogenic carbon preserved in a 4.1 billion-year-old
  zircon.
\newblock \emph{Proceedings of the National Academy of Sciences}, 112\penalty0
  (47):\penalty0 14518--14521, 2015.
\newblock \doi{10.1073/pnas.1517557112}.

\bibitem[{Bouvier}(2008)]{Bouvier:2008}
J.~{Bouvier}.
\newblock {Lithium depletion and the rotational history of exoplanet host
  stars}.
\newblock \emph{\aap}, 489:\penalty0 L53--L56, Oct. 2008.
\newblock \doi{10.1051/0004-6361:200810574}.

\bibitem[{Bressan} et~al.(2012){Bressan}, {Marigo}, {Girardi}, {Salasnich},
  {Dal Cero}, {Rubele}, and {Nanni}]{Bressan:2012}
A.~{Bressan}, P.~{Marigo}, L.~{Girardi}, B.~{Salasnich}, C.~{Dal Cero},
  S.~{Rubele}, and A.~{Nanni}.
\newblock {PARSEC: stellar tracks and isochrones with the PAdova and TRieste
  Stellar Evolution Code}.
\newblock \emph{\mnras}, 427:\penalty0 127--145, Nov. 2012.
\newblock \doi{10.1111/j.1365-2966.2012.21948.x}.

\bibitem[{Donahue} et~al.(1996){Donahue}, {Saar}, and {Baliunas}]{Donahue:1996}
R.~A. {Donahue}, S.~H. {Saar}, and S.~L. {Baliunas}.
\newblock {A Relationship between Mean Rotation Period in Lower Main-Sequence
  Stars and Its Observed Range}.
\newblock \emph{\apj}, 466:\penalty0 384, July 1996.
\newblock \doi{10.1086/177517}.

\bibitem[{Donahue} et~al.(1997){Donahue}, {Dobson}, and
  {Baliunas}]{Donahue:1997b}
R.~A. {Donahue}, A.~K. {Dobson}, and S.~L. {Baliunas}.
\newblock {Stellar Active Region Evolution - II. Identification and Evolution
  of Variance Morphologies in CA II H+K Time Series}.
\newblock \emph{\solphys}, 171:\penalty0 211--220, Mar. 1997.
\newblock \doi{10.1023/A:1004922323928}.

\bibitem[{Eddy}(1976)]{Eddy:1976}
J.~A. {Eddy}.
\newblock {The Maunder Minimum}.
\newblock \emph{Science}, 192:\penalty0 1189--1202, June 1976.
\newblock \doi{10.1126/science.192.4245.1189}.

\bibitem[{Egeland} et~al.(2015){Egeland}, {Metcalfe}, {Hall}, and
  {Henry}]{Egeland:2015}
R.~{Egeland}, T.~S. {Metcalfe}, J.~C. {Hall}, and G.~W. {Henry}.
\newblock {Sun-like Magnetic Cycles in the Rapidly-rotating Young Solar Analog
  HD 30495}.
\newblock \emph{\apj}, 812:\penalty0 12, Oct. 2015.
\newblock \doi{10.1088/0004-637X/812/1/12}.

\bibitem[{Egeland} et~al.(2016){Egeland}, {Soon}, {Baliunas}, {Hall},
  {Pevtsov}, and {Henry}]{Egeland:2016:cs19}
R.~{Egeland}, W.~{Soon}, S.~{Baliunas}, J.~C. {Hall}, A.~A. {Pevtsov}, and
  G.~W. {Henry}.
\newblock {Dynamo Sensitivity in Solar Analogs with 50 Years of Ca II H \& K
  Activity}.
\newblock In G.~A. Feiden, editor, \emph{Proceedings of the 19th Cambridge
  Workshop on Cool Stars, Stellar Systems, and the Sun}. Zenodo, Sept. 2016.
\newblock \doi{10.5281/zenodo.154118}.

\bibitem[Egeland et~al.(2016)Egeland, Soon, Baliunas, Hall, Pevtsov, and
  Henry]{Egeland:2016:cs19poster}
R.~Egeland, W.~Soon, S.~Baliunas, J.~C. Hall, A.~A. Pevtsov, and G.~W. Henry.
\newblock The solar dynamo zoo.
\newblock In \emph{The 19th Cambridge Workshop on Cool Stars, Stellar Systems,
  and the Sun}. Zenodo, 2016.
\newblock \doi{10.5281/zenodo.57920}.
\newblock URL \url{https://doi.org/10.5281/zenodo.57920}.

\bibitem[{Egeland} et~al.(2017){Egeland}, {Soon}, {Baliunas}, {Hall},
  {Pevtsov}, and {Bertello}]{Egeland:2017}
R.~{Egeland}, W.~{Soon}, S.~{Baliunas}, J.~C. {Hall}, A.~A. {Pevtsov}, and
  L.~{Bertello}.
\newblock {The Mount Wilson Observatory S-index of the Sun}.
\newblock \emph{\apj}, 835\penalty0 (1), January 2017.
\newblock \doi{10.3847/1538-4357/835/1/25}.

\bibitem[{Flower}(1996)]{Flower:1996}
P.~J. {Flower}.
\newblock {Transformations from Theoretical Hertzsprung-Russell Diagrams to
  Color-Magnitude Diagrams: Effective Temperatures, B-V Colors, and Bolometric
  Corrections}.
\newblock \emph{\apj}, 469:\penalty0 355, Sept. 1996.
\newblock \doi{10.1086/177785}.

\bibitem[{Gaidos} et~al.(2000){Gaidos}, {Henry}, and {Henry}]{Gaidos:2000}
E.~J. {Gaidos}, G.~W. {Henry}, and S.~M. {Henry}.
\newblock {Spectroscopy and Photometry of Nearby Young Solar Analogs}.
\newblock \emph{\aj}, 120:\penalty0 1006--1013, Aug. 2000.
\newblock \doi{10.1086/301488}.

\bibitem[{Gough}(1981)]{Gough:1981}
D.~O. {Gough}.
\newblock {Solar interior structure and luminosity variations}.
\newblock \emph{\solphys}, 74:\penalty0 21--34, Nov. 1981.
\newblock \doi{10.1007/BF00151270}.

\bibitem[{Hall} et~al.(2007){Hall}, {Lockwood}, and {Skiff}]{Hall:2007b}
J.~C. {Hall}, G.~W. {Lockwood}, and B.~A. {Skiff}.
\newblock {The Activity and Variability of the Sun and Sun-like Stars. I.
  Synoptic Ca II H and K Observations}.
\newblock \emph{\aj}, 133:\penalty0 862--881, Mar. 2007.
\newblock \doi{10.1086/510356}.

\bibitem[{Harder} et~al.(2009){Harder}, {Fontenla}, {Pilewskie}, {Richard}, and
  {Woods}]{Harder:2009}
J.~W. {Harder}, J.~M. {Fontenla}, P.~{Pilewskie}, E.~C. {Richard}, and T.~N.
  {Woods}.
\newblock {Trends in solar spectral irradiance variability in the visible and
  infrared}.
\newblock \emph{\grl}, 36:\penalty0 L07801, Apr. 2009.
\newblock \doi{10.1029/2008GL036797}.

\bibitem[{Harvey} and {White}(1999)]{Harvey:1999}
K.~L. {Harvey} and O.~R. {White}.
\newblock {Magnetic and Radiative Variability of Solar Surface Structures. I.
  Image Decomposition and Magnetic-Intensity Mapping}.
\newblock \emph{\apj}, 515:\penalty0 812--831, Apr. 1999.
\newblock \doi{10.1086/307035}.

\bibitem[{Henry} et~al.(1995){Henry}, {Fekel}, and {Hall}]{Henry:1995}
G.~W. {Henry}, F.~C. {Fekel}, and D.~S. {Hall}.
\newblock {An Automated Search for Variability in Chromospherically Active
  Stars}.
\newblock \emph{\aj}, 110:\penalty0 2926, Dec. 1995.
\newblock \doi{10.1086/117740}.

\bibitem[{Holmberg} et~al.(2009){Holmberg}, {Nordstr{\"o}m}, and
  {Andersen}]{Holmberg:2009}
J.~{Holmberg}, B.~{Nordstr{\"o}m}, and J.~{Andersen}.
\newblock {The Geneva-Copenhagen survey of the solar neighbourhood. III.
  Improved distances, ages, and kinematics}.
\newblock \emph{\aap}, 501:\penalty0 941--947, July 2009.
\newblock \doi{10.1051/0004-6361/200811191}.

\bibitem[{Judge} and {Egeland}(2015)]{Judge:2015}
P.~G. {Judge} and R.~{Egeland}.
\newblock {Century-long monitoring of solar irradiance and Earth's albedo using
  a stable scattering target in space}.
\newblock \emph{\mnras}, 448:\penalty0 L90--L93, Mar. 2015.
\newblock \doi{10.1093/mnrasl/slv004}.

\bibitem[{Kopp}(2014)]{Kopp:2014}
G.~{Kopp}.
\newblock {An assessment of the solar irradiance record for climate studies}.
\newblock \emph{Journal of Space Weather and Space Climate}, 4\penalty0
  (27):\penalty0 A14, Apr. 2014.
\newblock \doi{10.1051/swsc/2014012}.

\bibitem[{Lee} et~al.(2011){Lee}, {Beers}, {Allende Prieto}, {Lai}, {Rockosi},
  {Morrison}, {Johnson}, {An}, {Sivarani}, and {Yanny}]{Lee:2011}
Y.~S. {Lee}, T.~C. {Beers}, C.~{Allende Prieto}, D.~K. {Lai}, C.~M. {Rockosi},
  H.~L. {Morrison}, J.~A. {Johnson}, D.~{An}, T.~{Sivarani}, and B.~{Yanny}.
\newblock {The SEGUE Stellar Parameter Pipeline. V. Estimation of Alpha-element
  Abundance Ratios from Low-resolution SDSS/SEGUE Stellar Spectra}.
\newblock \emph{\aj}, 141:\penalty0 90, Mar. 2011.
\newblock \doi{10.1088/0004-6256/141/3/90}.

\bibitem[{Leighton}(1959)]{Leighton:1959}
R.~B. {Leighton}.
\newblock {Observations of Solar Magnetic Fields in Plage Regions.}
\newblock \emph{\apj}, 130:\penalty0 366, Sept. 1959.
\newblock \doi{10.1086/146727}.

\bibitem[{Li} et~al.(2012){Li}, {Bi}, {Liu}, {Tian}, and {Shuai}]{Li:2012}
T.~D. {Li}, S.~L. {Bi}, K.~{Liu}, Z.~J. {Tian}, and G.~Z. {Shuai}.
\newblock {Stellar parameters and seismological analysis of the star 18
  Scorpii}.
\newblock \emph{\aap}, 546:\penalty0 A83, Oct. 2012.
\newblock \doi{10.1051/0004-6361/201219063}.

\bibitem[{Linsky} and {Avrett}(1970)]{Linsky:1970}
J.~L. {Linsky} and E.~H. {Avrett}.
\newblock {The Solar H and K Lines}.
\newblock \emph{\pasp}, 82:\penalty0 169, Apr. 1970.
\newblock \doi{10.1086/128904}.

\bibitem[{Lockwood} et~al.(1997){Lockwood}, {Skiff}, and
  {Radick}]{Lockwood:1997}
G.~W. {Lockwood}, B.~A. {Skiff}, and R.~R. {Radick}.
\newblock {The Photometric Variability of Sun-like Stars: Observations and
  Results, 1984-1995}.
\newblock \emph{\apj}, 485:\penalty0 789--811, Aug. 1997.

\bibitem[{Lockwood} et~al.(2007){Lockwood}, {Skiff}, {Henry}, {Henry},
  {Radick}, {Baliunas}, {Donahue}, and {Soon}]{Lockwood:2007}
G.~W. {Lockwood}, B.~A. {Skiff}, G.~W. {Henry}, S.~{Henry}, R.~R. {Radick},
  S.~L. {Baliunas}, R.~A. {Donahue}, and W.~{Soon}.
\newblock {Patterns of Photometric and Chromospheric Variation among Sun-like
  Stars: A 20 Year Perspective}.
\newblock \emph{\apjs}, 171:\penalty0 260--303, July 2007.
\newblock \doi{10.1086/516752}.

\bibitem[{Meehl} et~al.(2013){Meehl}, {Arblaster}, and {Marsh}]{Meehl:2013}
G.~A. {Meehl}, J.~M. {Arblaster}, and D.~R. {Marsh}.
\newblock {Could a future ``Grand Solar Minimum'' like the Maunder Minimum stop
  global warming?}
\newblock \emph{\grl}, 40:\penalty0 1789--1793, May 2013.
\newblock \doi{10.1002/grl.50361}.

\bibitem[{Mel{\'e}ndez} et~al.(2014){Mel{\'e}ndez}, {Ram{\'{\i}}rez},
  {Karakas}, {Yong}, {Monroe}, {Bedell}, {Bergemann}, {Asplund}, {Tucci Maia},
  {Bean}, {do Nascimento}, {Bazot}, {Alves-Brito}, {Freitas}, and
  {Castro}]{Melendez:2014}
J.~{Mel{\'e}ndez}, I.~{Ram{\'{\i}}rez}, A.~I. {Karakas}, D.~{Yong}, T.~R.
  {Monroe}, M.~{Bedell}, M.~{Bergemann}, M.~{Asplund}, M.~{Tucci Maia},
  J.~{Bean}, J.-D. {do Nascimento}, Jr., M.~{Bazot}, A.~{Alves-Brito}, F.~C.
  {Freitas}, and M.~{Castro}.
\newblock {18 Sco: A Solar Twin Rich in Refractory and Neutron-capture
  Elements. Implications for Chemical Tagging}.
\newblock \emph{\apj}, 791:\penalty0 14, Aug. 2014.
\newblock \doi{10.1088/0004-637X/791/1/14}.

\bibitem[{Noyes} et~al.(1984){Noyes}, {Hartmann}, {Baliunas}, {Duncan}, and
  {Vaughan}]{Noyes:1984}
R.~W. {Noyes}, L.~W. {Hartmann}, S.~L. {Baliunas}, D.~K. {Duncan}, and A.~H.
  {Vaughan}.
\newblock {Rotation, convection, and magnetic activity in lower main-sequence
  stars}.
\newblock \emph{\apj}, 279:\penalty0 763--777, Apr. 1984.
\newblock \doi{10.1086/161945}.

\bibitem[{Petit} et~al.(2008){Petit}, {Dintrans}, {Solanki}, {Donati},
  {Auri{\`e}re}, {Ligni{\`e}res}, {Morin}, {Paletou}, {Ramirez Velez},
  {Catala}, and {Fares}]{Petit:2008}
P.~{Petit}, B.~{Dintrans}, S.~K. {Solanki}, J.-F. {Donati}, M.~{Auri{\`e}re},
  F.~{Ligni{\`e}res}, J.~{Morin}, F.~{Paletou}, J.~{Ramirez Velez},
  C.~{Catala}, and R.~{Fares}.
\newblock {Toroidal versus poloidal magnetic fields in Sun-like stars: a
  rotation threshold}.
\newblock \emph{\mnras}, 388:\penalty0 80--88, July 2008.
\newblock \doi{10.1111/j.1365-2966.2008.13411.x}.

\bibitem[{Pevtsov} et~al.(2016){Pevtsov}, {Virtanen}, {Mursula}, {Tlatov}, and
  {Bertello}]{Pevtsov:2016}
A.~A. {Pevtsov}, I.~{Virtanen}, K.~{Mursula}, A.~{Tlatov}, and L.~{Bertello}.
\newblock {Reconstructing solar magnetic fields from historical observations.
  I. Renormalized Ca K spectroheliograms and pseudo-magnetograms}.
\newblock \emph{\aap}, 585:\penalty0 A40, Jan. 2016.
\newblock \doi{10.1051/0004-6361/201526620}.

\bibitem[{Porto de Mello} and {da Silva}(1997)]{PortoDeMello:1997}
G.~F. {Porto de Mello} and L.~{da Silva}.
\newblock {HR 6060: The Closest Ever Solar Twin?}
\newblock \emph{\apjl}, 482:\penalty0 L89, June 1997.
\newblock \doi{10.1086/310693}.

\bibitem[{Pr{\v s}a} et~al.(2016){Pr{\v s}a}, {Harmanec}, {Torres}, {Mamajek},
  {Asplund}, {Capitaine}, {Christensen-Dalsgaard}, {Depagne}, {Haberreiter},
  {Hekker}, {Hilton}, {Kopp}, {Kostov}, {Kurtz}, {Laskar}, {Mason}, {Milone},
  {Montgomery}, {Richards}, {Schmutz}, {Schou}, and {Stewart}]{Prsa:2016}
A.~{Pr{\v s}a}, P.~{Harmanec}, G.~{Torres}, E.~{Mamajek}, M.~{Asplund},
  N.~{Capitaine}, J.~{Christensen-Dalsgaard}, {\'E}.~{Depagne},
  M.~{Haberreiter}, S.~{Hekker}, J.~{Hilton}, G.~{Kopp}, V.~{Kostov}, D.~W.
  {Kurtz}, J.~{Laskar}, B.~D. {Mason}, E.~F. {Milone}, M.~{Montgomery},
  M.~{Richards}, W.~{Schmutz}, J.~{Schou}, and S.~G. {Stewart}.
\newblock {Nominal Values for Selected Solar and Planetary Quantities: IAU 2015
  Resolution B3}.
\newblock \emph{\aj}, 152:\penalty0 41, Aug. 2016.
\newblock \doi{10.3847/0004-6256/152/2/41}.

\bibitem[{Radick} et~al.(1998){Radick}, {Lockwood}, {Skiff}, and
  {Baliunas}]{Radick:1998}
R.~R. {Radick}, G.~W. {Lockwood}, B.~A. {Skiff}, and S.~L. {Baliunas}.
\newblock {Patterns of Variation among Sun-like Stars}.
\newblock \emph{\apjs}, 118:\penalty0 239--258, Sept. 1998.
\newblock \doi{10.1086/313135}.

\bibitem[{Sagan} and {Mullen}(1972)]{Sagan:1972}
C.~{Sagan} and G.~{Mullen}.
\newblock {Earth and Mars: Evolution of Atmospheres and Surface Temperatures}.
\newblock \emph{Science}, 177:\penalty0 52--56, July 1972.
\newblock \doi{10.1126/science.177.4043.52}.

\bibitem[{Schwabe}(1844)]{Schwabe:1844}
M.~{Schwabe}.
\newblock {Sonnenbeobachtungen im Jahre 1843. Von Herrn Hofrath Schwabe in
  Dessau}.
\newblock \emph{Astronomische Nachrichten}, 21:\penalty0 233, Feb. 1844.

\bibitem[{Skumanich}(1972)]{Skumanich:1972}
A.~{Skumanich}.
\newblock {Time Scales for CA II Emission Decay, Rotational Braking, and
  Lithium Depletion}.
\newblock \emph{\apj}, 171:\penalty0 565, Feb. 1972.
\newblock \doi{10.1086/151310}.

\bibitem[{Solanki} et~al.(2013){Solanki}, {Krivova}, and {Haigh}]{Solanki:2013}
S.~K. {Solanki}, N.~A. {Krivova}, and J.~D. {Haigh}.
\newblock {Solar Irradiance Variability and Climate}.
\newblock \emph{\araa}, 51:\penalty0 311--351, Aug. 2013.
\newblock \doi{10.1146/annurev-astro-082812-141007}.

\bibitem[Stocker et~al.(2013)Stocker, Qin, Plattner, Tignor, Allen, Boschung,
  Nauels, Xia, Bex, and Midgley]{Stocker:2013}
T.~Stocker, D.~Qin, G.~Plattner, M.~Tignor, S.~Allen, J.~Boschung, A.~Nauels,
  Y.~Xia, B.~Bex, and B.~Midgley.
\newblock Ipcc, 2013: climate change 2013: the physical science basis.
  contribution of working group i to the fifth assessment report of the
  intergovernmental panel on climate change.
\newblock 2013.

\bibitem[{Willson} and {Hudson}(1991)]{Willson:1991}
R.~C. {Willson} and H.~S. {Hudson}.
\newblock {The sun's luminosity over a complete solar cycle}.
\newblock \emph{\nat}, 351:\penalty0 42--44, May 1991.
\newblock \doi{10.1038/351042a0}.

\bibitem[{Wilson}(1968)]{Wilson:1968}
O.~C. {Wilson}.
\newblock {Flux Measurements at the Centers of Stellar H- and K-Lines}.
\newblock \emph{\apj}, 153:\penalty0 221, July 1968.
\newblock \doi{10.1086/149652}.

\bibitem[{Wilson}(1978)]{Wilson:1978}
O.~C. {Wilson}.
\newblock {Chromospheric variations in main-sequence stars}.
\newblock \emph{\apj}, 226:\penalty0 379--396, Dec. 1978.
\newblock \doi{10.1086/156618}.

\bibitem[{Yeo} et~al.(2014){Yeo}, {Krivova}, and {Solanki}]{Yeo:2014}
K.~L. {Yeo}, N.~A. {Krivova}, and S.~K. {Solanki}.
\newblock {Solar Cycle Variation in Solar Irradiance}.
\newblock \emph{\ssr}, 186:\penalty0 137--167, Dec. 2014.
\newblock \doi{10.1007/s11214-014-0061-7}.

\end{thebibliography}

\end{document}